\documentclass[12pt]{article}
\usepackage{graphicx}
\usepackage{amsmath}
\usepackage{amssymb}
\usepackage{caption2}
\begin{document}
\bibliographystyle{prsty}
\begin{center}
{\large {\bf \sc{  Radiative  decays of the   $(0^+,1^+)$ strange-bottom  mesons}}} \\[2mm]
Zhi-Gang Wang \footnote{E-mail,wangzgyiti@yahoo.com.cn.  }    \\
Department of Physics, North China Electric Power University,
Baoding 071003, P. R. China
\end{center}

\begin{abstract}
In this article, we   assume that the  $(0^+,1^+)$ strange-bottom
mesons  are the conventional $b\bar{s}$ mesons, and calculate the
electromagnetic coupling constants $d$, $g_1$, $g_2$ and $g_3$ using
the light-cone QCD sum rules. Then we study the radiative decays
$B_{s0}\rightarrow B_s^* \gamma$, $B_{s1}\rightarrow B_s \gamma$,
$B_{s1}\rightarrow B_s^* \gamma$ and $B_{s1}\rightarrow B_{s0}
\gamma$, and observe that the widths are rather  narrow. We can
search for  the  $(0^+,1^+)$ strange-bottom  mesons in the invariant
$B_s \pi^0$ and $B^*_s \pi^0$ mass distributions in the strong
decays or in the invariant $B_s^*\gamma$ and $B_s\gamma$ mass
distributions in the radiative decays.
\end{abstract}


\section{Introduction}

 In 2007, the CDF Collaboration  reported the first observation of two narrow $B_s$ mesons
 with the spin-parity  $J^P=(1^+,2^+)$  using $1$
$\mathrm{fb^{-1}}$ of $p\overline{p}$ collisions at $\sqrt{s} = 1.96
\rm{TeV} $ collected with the CDF II detector at the Fermilab
Tevatron \cite{CDF}, the masses are $M_{B^*_{s1}}=( 5829.4 \pm 0.7)
\,\,\rm{MeV}$ and $M_{B_{s2}^*} = (5839.7 \pm 0.7)\,\,\rm{MeV}$. The
D0 Collaboration reported  the direct observation of the  $B_{s2}^*$
 in fully reconstructed decays to $B^+K^-$, the mass  is
 $(5839.6 \pm 1.1  \pm 0.7) \rm{MeV}$ \cite{D0}. While the  $(0^+,1^+)$ strange-bottom
 mesons   are still lack experimental evidence,
 they may be observed at the Tevatron or more probably  at the LHCb.
 The LHCb will be the most copious source of all the
$B$ hadrons, where the $b\bar{b}$  pairs will be copiously produced
with the cross section about $500 \,\mu b$ \cite{LHCbook}.

The $(0^+,1^+)$   doublet $B_s$ mesons  have been studied  with the
potential quark models, the heavy quark effective theory and the
lattice QCD
\cite{BBmeson1,BBmeson2,BBmeson3,BBmeson4,BBmeson5,BBmeson6,BBmeson7,BBmeson8,BBmeson9,
BBmeson10,Simonov07,Matsuki05,Matsuki07}, the predicted  masses are
different from each other.

 In the previous work \cite{Wang0712}, we have studied  the
 $(0^+,1^+)$ strange-bottom
 mesons using the QCD sum rules,  and observed   the central values
of the masses are below the corresponding $BK$ and $B^*K$
thresholds, respectively. It is a special property,
 the strong decays $B_{s0}\rightarrow BK$
and $B_{s1}\rightarrow B^*K$ are kinematically  forbidden. They can
decay through the isospin violation precesses $B_{s0}\rightarrow
B_s\eta\rightarrow B_s\pi^0$ and $B_{s1}\rightarrow
B_s^*\eta\rightarrow
 B_s^*\pi^0$ respectively, and the  widths are narrow
 \cite{Wang0801}. They can also decay through the
radiative processes.

Radiative decays are  important processes  in probing  the
structures of the hadrons  and  serve as valuable testing grounds to
select the best phenomenological model. The radiative decays $D^*\to
D\gamma$ and $B^*\to B\gamma$  have been studied extensively by
various theoretical approaches, such as the constituent quark model
\cite{GI,Barik,Jaus96,Ebert2002}, the light-cone QCD sum rules
\cite{Dosch96,Aliev96,ZHLi2002}, the heavy quark effective theory
\cite{Cheng1993,Colangelo1993},  the chiral perturbation theory
\cite{Amu1992}, the  light-front quark model \cite{Choi2007}, etc
(For more literatures, one can consult the excellent review
"Phenomenology of heavy meson chiral Lagrangians" \cite{HQET1997} or
the book "Heavy quark physics"\cite{HQPbook}).   The works on the
radiative decays of the $P$-wave heavy-light
 mesons are relatively
fewer and focus on the charm mesons $D_{s0}$ and $D_{s1}$.  The
radiative decays  of the $D_{s0}$ and $D_{s1}$ have been studied
using the constituent quark model
\cite{BBmeson3,Godfrey2003PLB,Liu2006CQM,Godfrey2005,Close2005CQM},
the vector meson dominance ansatz in the heavy quark limit
\cite{Colangelo2003PLB,Colangelo2004MPLA},   the heavy-hadron chiral
perturbation theory \cite{Mehen2004},  the light-cone QCD sum rules
\cite{RadiativeD,Wang2007R}, the  effective $SU(4)$ theory with
dynamically generated  scalar resonances \cite{Gamermann2007}, etc.

In Ref.\cite{RadiativeD}, the radiative decays $D_{s0}\rightarrow
D_s^* \gamma$, $D_{s1}\rightarrow D_s \gamma$, $D_{s1}\rightarrow
D_s^* \gamma$ and $D_{s1}\rightarrow D_{s0} \gamma$ are studied
using the light-cone QCD sum rules. Experimentally, the branching
fractions listed in the particle data group are $\rm{Br}(D_{s1}\to
D_s \gamma)=(18\pm 4)\%$, $\rm{Br}(D_{s1}\to D^*_s \gamma)<8\%$ and
$\rm{Br}(D_{s1}\to D_{s0} \gamma)=3.7^{+5.1}_{-2.4}\%$ \cite{PDG}.

The $(0^+,1^+)$  $D_s$ and $B_s$ mesons have similar properties
\cite{Wang0712,Wang0801,Wang08}, we extend our previous works  to
make systematic studies. The mesons $B_{s0}$ and $B_{s1}$ may be
observed in the invariant $B_s^*\gamma$ and $B_s\gamma$ mass
distributions, and the radiative decays  are suitable to understand
the nature of the strange-bottom mesons. In this article, we  study
the radiative decays $B_{s0}\rightarrow B_s^* \gamma$,
$B_{s1}\rightarrow B_s \gamma$, $B_{s1}\rightarrow B_s^* \gamma$ and
$B_{s1}\rightarrow B_{s0} \gamma$ using the light-cone QCD sum
rules.

The neutral strange-bottom mesons and charged strange-charm mesons
have different electromagnetic properties besides   they have
different masses. The present work is far from trivial as just a
replacement $c \to b$, we can borrow some ideas from the magnetic
moments of the nucleons.

In the isospin limit, the proton and neutron have degenerated mass,
however, their electromagnetic properties are quite different. If we
take them as point particles, their magnetic moments are $\mu_p=1$
and $\mu_n=0$ (in unit of nucleon magneton) from  Dirac's theory of
relativistic fermions. In 1933, Otto Stern measured the magnetic
moment of the proton, which deviates from one significatively and
indicates the proton has under-structures.  The neutron is neutral,
its (anomalous) magnetic  originates  from the Pauli form-factor.
The electromagnetic form-factors (Dirac and Pauli form-factor) are
excellent subjects to under the under-structures of the nucleon, and
have been  extensively studied both experimentally and
theoretically.

The radiative decays embody  the nature of the hadron's constituents
and the dynamics that binds the constituents together, the present
work is necessary.

The light-cone QCD sum rules approach carries out the operator
product expansion near the light-cone $x^2\approx 0$ instead of the
short distance $x\approx 0$ while the non-perturbative matrix
elements are parameterized by the light-cone distribution amplitudes
   instead of
 the vacuum condensates \cite{LCSR1,LCSR2,LCSR3,LCSR4,LCSR5,LCSRreview}. The
 coefficients in the light-cone distribution amplitudes are calculated
 with the conventional QCD  sum rules
 and the  values are universal \cite{SVZ791,SVZ792,Reinders85}.

The article is arranged as: in Section 2, we derive the
electromagnetic  coupling constants  $d$, $g_1$, $g_2$ and $g_3$
using  the light-cone QCD sum rules; in Section 3, the numerical
result and discussion; and Section 4 is reserved for conclusion.

\section{ Electromagnetic  coupling constants  $d$, $g_1$, $g_2$ and $g_3$ with light-cone QCD sum rules}

In the following, we write down the definitions  for the
electromagnetic coupling constants ($d$, $g_1$, $g_2$ and $g_3$)
among the  $(0^-,1^-)$, $(0^+,1^+)$ strange-bottom  mesons and the
photon \cite{RadiativeD},
\begin{eqnarray}
 \langle \gamma(q) B_s^*(p)|
B_{s0}(k)\rangle &=&  e d  \left\{\varepsilon^* \cdot \tilde \eta^*
p\cdot q-\varepsilon^* \cdot p \tilde \eta^* \cdot q
\right\}  \,  , \nonumber\\
\langle \gamma(q) B_s(p)| B_{s1}(k)\rangle &=& e g_1  \left\{
\varepsilon^* \cdot  \eta p \cdot q - \varepsilon^* \cdot p \eta
\cdot q  \right\}\ , \nonumber\\
\langle \gamma(q) B^*_s(p)| B_{s1}(k)\rangle &=& i \, e \, g_2 \,
\varepsilon_{\alpha \beta \sigma \tau} \eta^\alpha \tilde
\eta^{*\beta} \varepsilon^{*\sigma}   q^\tau  \,  ,\nonumber\\
\langle \gamma(q) B_{s0} (p)| B_{s1}(k)\rangle& =& i \, e \,  g_3 \,
\varepsilon_{\alpha \beta \sigma \tau}  \varepsilon^{*\alpha}
\eta^\beta    p^\sigma q^\tau \, ,
\end{eqnarray}
where the $\varepsilon$, $\eta$ and $\tilde \eta$  are the
polarization vectors of  the  photon,  $B_{s1}$ and $B_s^*$
respectively, and $e$ is the electric charge. In
Ref.\cite{RadiativeD}, the radiative decays of the $(0^+,1^+)$
strange-charm  mesons are studied using the light-cone QCD sum
rules, in this article, we follow the routine and study the
radiative decays of the  $(0^+,1^+)$ strange-bottom  mesons.

We study the electromagnetic  coupling constants  $d$, $g_1$, $g_2$
and $g_3$ with the two-point correlation functions $F_{\mu}(p,q)$,
$T_{\mu}(p,q)$, $T_{\mu\nu}(p,q)$ and $W_{\mu}(p,q)$, respectively,
\begin{eqnarray}
 F_\mu(p,q)&=&i\int d^4x \; e^{i p \cdot x} \langle \gamma(q) |
T\left\{J^\dagger_\mu(x) J_0(0)\right\} |0\rangle \, ,\nonumber \\
T_\mu(p,q)&=&i \int d^4x  e^{i p \cdot x} \langle
\gamma(q) | T\left\{J^\dagger_5(x) J^A_\mu(0)\right\}|0\rangle \, ,\nonumber \\
T_{\mu \nu}(p,q)&=&i \int d^4x  e^{i p \cdot x} \langle
\gamma(q) | T\left\{ J^\dagger_\mu(x) J^A_\nu(0) \right\} |0\rangle \, ,\nonumber \\
W_\mu(p,q)&=&i \int d^4x \ e^{i p \cdot x} \langle \gamma(q) |
T\left\{J^\dagger_0(x) J^A_\mu(0)\right\} |0\rangle \, ,
\end{eqnarray}
where
\begin{eqnarray}
J_0(x)&=&\bar{b}(x) s(x) \, , \nonumber\\
J_\mu(x)&=&\bar{b}(x)\gamma_\mu s(x) \, , \nonumber\\
J^A_\mu(x)&=&\bar{b}(x)\gamma_\mu \gamma_5 s(x) \, , \nonumber\\
J_5(x)&=&\bar{b}(x)i\gamma_5 s(x) \, .
\end{eqnarray}
The current operators $J_0(x)$, $J_5(x)$, $J_\mu(x)$ and
$J^A_\mu(x)$ interpolate the mesons $B_{s0}$, $B_s$, $B_s^*$ and
$B_{s1}$ respectively. The correlation functions $F_{\mu}(p,q)$,
$T_{\mu}(p,q)$, $T_{\mu\nu}(p,q)$ and $W_{\mu}(p,q)$ can be
decomposed as
\begin{eqnarray}
 F_\mu(p,q)&=& F_A  \left(
p \cdot \varepsilon^*  q_\mu - p\cdot q \varepsilon^* _\mu
\right) + \cdots \, , \nonumber \\
T_\mu(p,q)&=& T_B    \left(  p\cdot\varepsilon^*    q_\mu -
p\cdot q \varepsilon^* _\mu  \right)  + \cdots \, ,\nonumber \\
T_{\mu \nu}(p,q)&=& T_C     \varepsilon_{\mu \nu \sigma \tau}
\varepsilon^{* \sigma} q^\tau +
T_1    p_\mu  \varepsilon_{\nu \beta \sigma \tau} p^\beta \varepsilon^{* \sigma} q^\tau \nonumber\\
&&+T_2    (p+q)_\nu \varepsilon_{\alpha \mu  \sigma \tau} p^\alpha
\varepsilon^{* \sigma} q^\tau + \cdots \, , \nonumber \\
W_\mu (p,q)&=&  W_D i  \varepsilon_{\mu \alpha  \sigma \tau}
\varepsilon^{* \alpha} p^\sigma q^\tau
\end{eqnarray}
due to  Lorentz invariance. We choose the tensor structures $ p
\cdot \varepsilon^*  q_\mu - p\cdot q \varepsilon^* _\mu $,
$\varepsilon_{\mu \nu \sigma \tau} \varepsilon^{* \sigma} q^\tau$
and $\varepsilon_{\mu \alpha  \sigma \tau} \varepsilon^{* \alpha}
p^\sigma q^\tau$ for analysis.

In this article, we consult the analytical expressions of
Ref.\cite{RadiativeD}, and make a simple replacement for  the
corresponding  parameters of the strange-charm  and strange-bottom
mesons to obtain the following four sum rules. We would like not to
follow  the standard procedure of the light-cone QCD sum rules and
repeat the straightforward but tedious calculations, one can consult
Ref.\cite{RadiativeD} for the technical details. We perform detailed
numerical calculations, analyze the effects originate from the
electric charge difference between the $c$ and $b$ quarks in
addition to the heavy quark symmetry. Taking into account our
previous works \cite{Wang0712,Wang0801,Wang08}, we  make systematic
studies about the properties of the $(0^+,1^+)$ $B_s$ mesons.
\begin{eqnarray}
d&=& \frac{\exp\left( \frac{M^2_{B_{s0}}+M^2_{B^*_s}}{ 2 M^2 }\right)}{ f_{B_{s0}} f_{B^*_s} M_{B_{s0}} M_{B^*_s} }
\left\{ \int_{\Delta}^{s^0_A} d s   e^{- {s \over M^2}} \rho_A(s)- 2 e_s f_{3 \gamma} m_b e^{-{m_b^2 \over M^2}}
\Psi^v(u_0) \right. \nonumber\\
&& + e_b  e^{-{m_b^2 \over M^2}} \langle\bar{s}s\rangle
\left(1+{m_s^2 \over 4 M^2} +{m_s^2 m_b^2 \over 2 M^4} \right)
+ e_s \langle\bar{s}s\rangle (e^{-{m_b^2 \over M^2}}- e^{-{s^0_A \over M^2}}) M^2 \chi \phi_\gamma(u_0) \nonumber\\
&& +e_s \langle\bar{s}s\rangle e^{-{m_b^2 \over M^2}}  \left[ -\frac{1}{4} \left[\mathbb{A}(u_0)-8 \bar H_\gamma(u_0)\right]\left(1+{m_b^2\over M^2}\right) \right.\nonumber\\
&&+\int_0^{1-u_0} dv \int_0^{u_0 \over 1-v} d \alpha_g  {\cal F}_A(u_0-(1-v)\alpha_g, 1-u_0-v \alpha_g,\alpha_g)\nonumber\\
&&\left.\left.+\int^1_{1-u_0} d v \int_0^{1-u_0 \over v} d \alpha_g
{\cal F}_A(u_0-(1-v)\alpha_g, 1-u_0-v
\alpha_g,\alpha_g)\right]\right\}\, ,
\end{eqnarray}

\begin{eqnarray}
g_1&=& \frac{\exp\left(\frac{M^2_{B_{s1}}+M^2_{B_s}}{ 2 M^2 }\right)
(m_b+m_s)}{ f_{B_{s1}}f_{B_s} M^2_{B_s} M_{B_{s1}}} \left\{
\int_{\Delta}^{s_B^0} d s   e^{- {s \over M^2}} \rho_B(s)
+ 2 e_s f_{3 \gamma} m_b e^{-{m_b^2 \over M^2}}  \Psi^v(u_0) \right.\nonumber\\
&& +e_b  e^{-{m_b^2 \over M^2}} \langle\bar{s}s\rangle \left[1-{m_b
m_s \over M^2} +{m_s^2 \over 2 M^2}
\left(1+{m_b^2 \over M^2}\right) \right]   \nonumber\\
&&- e_s \langle\bar{s}s\rangle (e^{-{m_b^2 \over M^2}}- e^{-{s^0_B \over M^2}}) M^2 \chi \phi_\gamma(u_0) \nonumber\\
&& -e_s \langle\bar{s}s\rangle e^{-{m_b^2 \over M^2}} \left[ -\frac{1}{4} \left[\mathbb{A}(u_0)-8 \bar H_\gamma(u_0)\right]\left(1+{m_b^2\over M^2}\right)\right.\nonumber\\
&&-\int_0^{1-u_0} dv \int_0^{u_0 \over 1-v} d \alpha_g  {\cal F}_B(u_0-(1-v)\alpha_g, 1-u_0-v \alpha_g,\alpha_g)\nonumber\\
&&\left.\left.-\int^1_{1-u_0} d v \int_0^{1-u_0 \over v} d \alpha_g
{\cal F}_B(u_0-(1-v)\alpha_g, 1-u_0-v
\alpha_g,\alpha_g)\right]\right\}\, ,
\end{eqnarray}

\begin{eqnarray}
 g_2&=& \frac{\exp\left(\frac{M^2_{B_{s1}}+M^2_{B^*_s}}{ 2 M^2
}\right)}{ f_{B_{s1}} f_{B^*_s} M_{B_{s1}}M_{B^*_s} } \left\{
\int_{\Delta}^{s^0_C} d s   e^{- {s \over M^2}} \rho_C(s)
+ e_b  m_b  e^{-{m_b^2 \over M^2}} \langle\bar{s}s\rangle \left[1-{m_s^2 \over  M^2}+{m_s^2m_b^2 \over M^4}  \right]  \right.  \nonumber\\
&& +e_s m_b \langle\bar{s}s\rangle  (e^{-{m_b^2 \over M^2}}- e^{-{s^0_C \over M^2}}) M^2 \chi \phi_\gamma(u_0)  \nonumber\\
&&+ e_s m_b \langle\bar{s}s\rangle  e^{-{m_b^2 \over M^2}} \left[
-\frac{1}{4} {m_b^2 \over M^2}
\mathbb{A}(u_0) -H_\gamma(u_0)(1-u_0)- \bar H_\gamma(u_0) \left(1- \frac{2 m_b^2}{M^2}\right)\right]\nonumber\\
&&+ e_s f_{3 \gamma}  M^2 (e^{-{m_b^2 \over M^2}} - e^{-{s^0_C \over
M^2}}) \left[ \frac{1}{4} (1-u_0) \psi^{\prime a} (u_0) -
 \frac{1}{4}  \psi^a (u_0)\right.\nonumber\\
 &&\left.-  \Psi^v (u_0) \left(1+\frac{2 m_b^2}{M^2}\right)+(1-u_0)   \psi^v (u_0) \right] \nonumber\\
 && +m_b e_s \langle\bar{s}s\rangle  e^{-{m_b^2 \over M^2}} \Bigg[
 \int_0^{1-u_0} dv \int_0^{u_0 \over 1-v} d \alpha_g  {\cal F}_{C1}
 (u_0-(1-v)\alpha_g, 1-u_0-v \alpha_g,\alpha_g)\nonumber\\
&&+\int^1_{1-u_0} d v \int_0^{1-u_0 \over v} d \alpha_g  {\cal F}_{C1}
(u_0-(1-v)\alpha_g, 1-u_0-v \alpha_g,\alpha_g)\Bigg]\nonumber\\
&& - e_s f_{3 \gamma} M^2 (e^{-{m_b^2 \over M^2}} - e^{-{s^0_C \over
M^2}}) \left[
 \int_0^{u_0} d \alpha_{\bar q}    \int_{u_0-\alpha_{\bar q}}^{1-
 \alpha_{\bar q}}  \frac{d \alpha_g}{\alpha_g^2} {\cal F}_{C2} (1- \alpha_{\bar q}
 -\alpha_g, \alpha_{\bar q} ,\alpha_g)\right.\nonumber\\
&&\left.\left.-\int_0^{u_0} d  \alpha_{\bar q}
\frac{1}{u_0-\alpha_{\bar q}}
 {\cal F}_{C2} (1-u_0,  \alpha_{\bar q},u_0-\alpha_{\bar q})
\right]\right\} \, ,
\end{eqnarray}

\begin{eqnarray}
g_3&=& \frac{\exp\left( \frac{ M^2_{B_{s1}}+M^2_{B_{s0}} }{ 2 M^2 }
\right)}{ f_{B_{s1}}f_{B_{s0}}  M_{B_{s1}}M_{B_{s0}} } \left\{
\int_{\Delta}^{s_D^0} d s    e^{- {s \over M^2}} \rho_D(s)
+ e_b e^{-{m_b^2 \over M^2}}  \langle\bar{s}s\rangle \left(1+{m_s m_b \over 2 M^2}+{m_s^2 m_b^2 \over 8 M^4}\right)  \right.\nonumber\\
&&+ e_s \langle\bar{s}s\rangle (e^{-{m_b^2 \over M^2}} - e^{-{s^0_C
\over M^2}} )
M^2 \chi \phi_\gamma(u_0) \nonumber\\
&& +e^{-{m_b^2 \over M^2}}  e_s \langle\bar{s}s\rangle [- \frac{1}{4} \mathbb{A}(u_0)
 (1+\frac{m_b^2}{M^2})]-\frac{m_b}{2} e_s f_{3 \gamma} \psi^a(u_0)  e^{-{m_b^2 \over M^2}}  \nonumber\\
&&+e^{-{m_b^2 \over M^2}} e_s \langle\bar{s}s\rangle
\left[\int^{1-u_0}_0 d v \int_0^{u_0 \over 1- v} d \alpha_g  {\cal F}_D(u_0-(1-v)\alpha_g, 1-u_0-v \alpha_g,\alpha_g) \right.\nonumber\\
&& \left.\left.+\int^1_{1-u_0} d v \int_0^{1-u_0 \over v} d \alpha_g
{\cal F}_D(u_0-(1-v)\alpha_g, 1-u_0-v \alpha_g,\alpha_g) \right]
 \right\} \, ,
 \end{eqnarray}
where
\begin{eqnarray}
\rho_A(s)&=& \frac{3 e_s}{4 \pi^2} \left\{m_s \ln \left({s-m_b^2+m_s^2-\lambda^\frac{1}{2}(s, m_b^2,m_s^2)
\over s-m_b^2+m_s^2+\lambda^{1\over 2}(s, m_b^2,m_s^2) } \right) -{m_b-m_s \over s}
 \lambda^{1\over 2}(s, m_b^2,m_s^2) \right\} \nonumber\\
&&+{3 e_s \over 4 \pi^2} {m_b+m_s \over 2}  {\lambda^{1\over 2}(s,
m_b^2,m_s^2) \over s} \left(1-{m_s^2-m_b^2 \over s } \right)  +
(s \leftrightarrow b) \, ,\nonumber\\
\rho_B(s)&=&- {3 e_s \over 8 \pi^2} \left\{ 2 m_s \ln \left({s-m_b^2+m_s^2-\lambda^{1\over 2}(s, m_b^2,m_s^2) \over s-m_b^2+m_s^2+\lambda^{1\over 2}(s, m_b^2,m_s^2) } \right)\right. \nonumber\\
&&\left.+(m_b-m_s){(m_b^2-m_s^2-s) \over s^2}  \lambda^{1\over 2}(s,
m_b^2,m_s^2) \right\}  - (s \leftrightarrow b) \, , \nonumber\\
\rho_C(s)&=& {3 e_s \over 4 \pi^2}  m_s m_b \ln
\left({s-m_b^2+m_s^2-\lambda^{1\over 2}(s, m_b^2,m_s^2) \over
s-m_b^2+m_s^2+\lambda^{1\over 2}(s, m_b^2,m_s^2) } \right)
   +  (s \leftrightarrow b) \,  , \nonumber\\
 \rho_D(s)&=& {3 e_s \over 4 \pi^2} \left\{ \frac{m_b+m_s}{s} \lambda^{1\over 2}(s, m_b^2,m_s^2)  + m_s \ln
 \left({s-m_b^2+m_s^2-\lambda^{1\over 2}(s, m_b^2,m_s^2) \over s-m_b^2+m_s^2+\lambda^{1\over 2}(s, m_b^2,m_s^2) } \right) \right\} \nonumber\\
&&- (s \leftrightarrow b) \,  , \nonumber\\
{\cal F}_A&=& {\cal S} - \tilde{\cal S}-T_1+T_4-T_3+T_2+2 v (- {\cal
S}+T_3-T_2) \, , \nonumber\\
{\cal F}_B &=& {\cal S} + \tilde{\cal S}-T_1-T_2+T_3+T_4+2 v (-
{\cal S}-T_3+T_2)\, , \nonumber\\
{\cal F}_{C1}&=& {\cal S} + \tilde{\cal S}+T_1-T_2-T_3+T_4 \, ,\nonumber\\
{\cal F}_{C2}&=& {\cal A} + {\cal V} \, ,\nonumber\\
{\cal F}_D &=& {\cal S} + \tilde{\cal S}+T_1+T_4-T_2-T_3+2 v (-
\tilde{\cal S}+T_3-T_4) \, ,
\end{eqnarray}
and $\Delta= (m_b+m_s)^2$, $ \bar H_\gamma(u)=\int_0^u d u^\prime
H_\gamma(u^\prime)$, $ H_\gamma(u)=\int_0^u d u^\prime
h_\gamma(u^\prime)$, $ \Psi^v(u)=\int_0^u d u^\prime
\psi^v(u^\prime)$. The explicit expressions  of the light-cone
distribution amplitudes ${\cal S}$, $\tilde{\cal S}$, $T_1$, $T_2$,
$T_3$, $T_4$, ${\cal A}$, ${\cal V}$, $\mathbb{A}$, $\phi_\gamma$,
$\psi^a$, $\psi^v$ and $h_\gamma$ are given in the appendix
\cite{LCDA}. In Eqs.(5-8), the standard definitions for the decay
constants have been used,
\begin{eqnarray}
\langle0 | J^\dagger_{\mu}(0)|B_s^*(p)\rangle&=&f_{B^*_s}M_{B^*_s}\tilde{ \eta}_\mu\,, \nonumber\\
\langle0|J^{A\dagger}_{\mu}(0)|B_{s1}(p)\rangle&=&f_{B_{s1}}M_{B_{s1}}\eta_\mu\,,\nonumber \\
\langle0 | J^\dagger_5(0)|B_s(p)\rangle&=&\frac{f_{B_s}M^2_{B_s}}{m_b+m_s}\,, \nonumber\\
\langle0 | J^\dagger_0(0)|B_{s0}(p)\rangle&=&f_{B_{s0}}M_{B_{s0}} \,
.
 \end{eqnarray}

The parameters in the light-cone distribution amplitudes are scale
dependent and  calculated  using  the QCD sum rules \cite{LCDA}. In
the heavy quark limit, the bound energy of the $(0^+,1^+)$
strange-bottom  mesons is about
$\Lambda=\frac{3M_{B_{s1}}+M_{B_{s0}}}{4}-m_b\approx 1\,\,\rm{GeV}$,
which can serve as a typical energy scale and validate our choice
$\mu=1\,\,\rm{GeV}$,  one can choose another typical energy scale
$\mu=\sqrt{M_B^2-m_b^2}\approx 2.4\,\,\rm{GeV}$. The physical
quantities would not depend on the special energy scale we choose,
we expect that scale dependence of the input parameters is canceled
out approximately with each other, the values of the electromagnetic
coupling constants which calculated  at the energy scale
$\mu=1\,\,\rm{GeV}$ can make robust predictions.

The masses of the strange-bottom mesons are
$M_{B_{s1}}=5.72\,\,\rm{GeV}$, $M_{B_{s0}}=5.70\,\,\rm{GeV}$,
$M_{B^*_s}=5.412\,\,\rm{GeV}$ and $M_{B_s}=5.366\,\,\rm{GeV}$,
\begin{eqnarray}
 \frac{M_{B_{s0}}^2}{M_{B_{s0}}^2+M_{B_s^*}^2}\approx\frac{M_{B_{s1}}^2}{M_{B_{s1}}^2+M_{B_s}^2}\approx\frac{M_{B_{s1}}^2}{M_{B_{s1}}^2+M_{B_s^*}^2}\approx
\frac{M_{B_{s1}}^2}{M_{B_{s0}}^2+M_{B_{s1}}^2}\approx 0.50-0.53 \, .
\nonumber
\end{eqnarray}
There exists an overlapping working window for the two Borel
parameters $M_1^2$ and $M_2^2$. It is convenient to take the value
$M_1^2=M_2^2=2M^2$ and $u_0=\frac{1}{2}$.

In the four sum rules, the terms originate  from the nonperturbative
interactions between the photon and  quarks  can be classified as
$\mathcal {O}(M^2)$, $\mathcal {O}(1)$, $\mathcal
{O}(\frac{1}{M^2})$, $\cdots$, the terms of order $\mathcal
{O}(M^2)$ are greatly enhanced by the large Borel parameter $M^2$,
their  contributions are large and continuum subtraction is
necessary.   We introduce the threshold parameter $s_0$ (denotes
$s_A^0$, $s_B^0$, $s_C^0$ and $s_D^0$) and make the simple
replacement $e^{-\frac{m_b^2}{M^2}} \rightarrow
e^{-\frac{m_b^2}{M^2} }-e^{-\frac{s_0}{M^2}} $ for the terms of
order $\mathcal {O}(M^2)$ to subtract the contaminations from the
high resonances  and
  continuum states. For technical details, one can consult Ref.\cite{Belyaev94}.

The $(0^+,1^+)$ $D_s$ and $B_s$ mesons may have
 $c\bar{s}$ and $b\bar{s}$ kernels of the typical
 $c\bar{s}$ and $b\bar{s}$  mesons size respectively,
strong couplings to the virtual intermediate hadronic states (or
virtual mesons loops) may result in smaller masses than the
conventional   $c\bar{s}$ and $b\bar{s}$ mesons in the potential
 models
\cite{Wang08,WangD1,WangD2,Wang06}. In Ref.\cite{Guo0712}, Guo et al
 take the masses from the potential models as  bare  masses,
 and calculate the mass shifts for the scalar  heavy  mesons
 due to the hadronic loops, the numerical  results indicate   the masses
from the  quark models can be reduced significantly.
    In the previous works, we have calculated the strong coupling constants
 $g_{D_{s0} DK}$,
$g_{D_{s1} D^*K}$, $g_{B_{s0} BK}$ and $g_{B_{s1} B^*K}$  using the
light-cone QCD sum rules \cite{Wang08,WangD1,WangD2,Wang06}, the
large strong coupling constants support the hadronic dressing
mechanism \cite{HDress1,HDress2,HDress3}. In this article, we assume
that the hadronic loops reduce "bare" masses from the potential
models, not the (renormalized) physical masses from the QCD sum
rules, and neglect possible contaminations from the $BK$ and $B^*K$
thresholds.

\section{Numerical result and discussion}
The input parameters    are taken as $f_{3
\gamma}=-(0.0039\pm0.0020)\,\, {\rm GeV}^2$, $\omega^V_\gamma =
3.8\pm 1.8$, $\omega^A_\gamma = -2.1\pm  1.0$, $\chi=-(3.15 \pm0.3)
\,\,{\rm GeV}^{-2}$ \cite{LCDA}, $k=0.2$, $\zeta_1=0.4$,
$\zeta_2=0.3$, $\varphi_2=k^+=\zeta_1^+=\zeta_2^+=0$ \cite{LCSR1},
$\langle \bar s s \rangle=0.8 \langle \bar q q \rangle$, $\langle
\bar q q \rangle=(-0.24 \,\,\rm{ GeV})^3 $
\cite{SVZ791,SVZ792,Reinders85},
 $m_s=(0.14\pm 0.01 )\,\,\rm{GeV}$,
 $m_b=(4.7\pm 0.1)\,\,\rm{GeV}$, $M_{B_s}=5.366\,\,\rm{GeV}$,
$M_{B^*_s}=5.412\,\,\rm{GeV}$ \cite{PDG},
$M_{B_{s0}}=5.70\,\,\rm{GeV} $, $M_{B_{s1}}=5.72\,\,\rm{GeV}$,
$f_{B_{s0}}= f_{B_{s1}}=0.24\,\,\rm{GeV}$ \cite{Wang0712}, and
$f_{B^*_s}=f_{B_s}=0.19\,\,\rm{GeV}$
\cite{LCSRreview,Wang04BS,Verde-Velasco07}.

The threshold parameters are taken as $s_S^0=(37\pm1)\,\,\rm{GeV}^2$
and $s^0_A=(38\pm1)\,\,\rm{GeV}^2$, which are chosen to below the
corresponding masses of the first radially  excited states,
$M_{Sr}=6.264\,\,\rm{GeV}$ for the $B_{s0}$ and
$M_{Ar}=6.296\,\,\rm{GeV}$ for the $B_{s1} $  in the potential
models \cite{BBmeson6}.

In 2006, the BaBar Collaboration observed a new $c\bar{s}$ state
$D_s(2860)$ with the mass $M=(2856.6 \pm 1.5 \pm 5.0)\,\,\rm{MeV}$,
width  $\Gamma=(48 \pm 7 \pm 10)\,\,\rm{MeV}$ and possible
spin-parity $0^+$, $1^-$, $2^+, \cdots$ \cite{BaBar0607}. It has
been interpreted as the first radial excitation of the $0^+$ state
$D_{s0}(2317)$ in Refs.\cite{2Ds-1,2Ds-2}, although other
identifications  are not excluded. The energy gap between the $2P$
and $1P$ scalar $c\bar{s}$ states is about $\delta
M_S=0.539\,\,\rm{GeV}$.

If the masses of the $P$-wave strange-bottom mesons are of the same
order (about $5.8\,\rm{GeV}$ \cite{CDF,D0}) and the energy gap
between the ground state and the first radially  excited state is
about $0.5\,\,\rm{GeV}$ (just like the $c\bar{s}$ mesons), we can
make a rough estimation for the masses of the first radially excited
$(0^+,1^+)$ strange-bottom states, $M_r\approx
(5.8+0.5)\,\,\rm{GeV}$. The threshold parameters should be chosen as
$s_0< M_r^2\approx40\,\,\rm{GeV}^2$, which are consistent with the
predictions of the potential models \cite{BBmeson6}.

 The Borel parameters are chosen as $M^2=(5-7)\,\,\rm{GeV}^2$,
   which are determined from the two-point QCD sum rules \cite{Wang0712}.
  In those regions, the
contributions from the pole terms are larger than $50\%$,
furthermore, the dominating contributions come from the perturbative
terms.

The  masses $M_{B_{s0}}$ and $M_{B_{s1}}$ obtained from the QCD sum
rules have uncertainties, $M_{B_{s0}}=(5.70\pm0.11)\,\,\rm{GeV} $
and $M_{B_{s1}}=(5.72\pm0.09)\,\,\rm{GeV}$ \cite{Wang0712},  we can
take the central values to avoid the possibility $M_{B_{s0}}
>M_{B_{s1}} $, in that case the radiative decay $B_{s1}\rightarrow
B_{s0} \gamma$ is kinematically forbidden. Furthermore, we neglect
the uncertainties of the  decay constants for consistence. The
masses and decay constants of the $(0^+,1^+)$ mesons  are calculated
using the QCD sum rules, some uncertainties originate from the Borel
parameters and threshold parameters, so our approximation is not
crude.

Taking into account  the uncertainties of the input parameters,
finally we obtain the numerical values of the electromagnetic
coupling constants $d$, $g_1$, $g_2$, and $g_3$ (which are shown in
Figs.(1-4) respectively)
\begin{eqnarray}
  |d| &=&(0.09-0.29) \,\,\rm{GeV}^{-1} \, , \nonumber\\
  |g_1| &=&(0.18-0.40)\,\, \rm{GeV}^{-1} \, , \nonumber\\
  |g_2| &=&0.25-1.20 \, , \nonumber\\
  |g_3| &=&(0.31-0.64) \,\,\rm{GeV}^{-1} \, ,
 \end{eqnarray}
\begin{figure} \centering
  \includegraphics[totalheight=8cm,width=10cm]{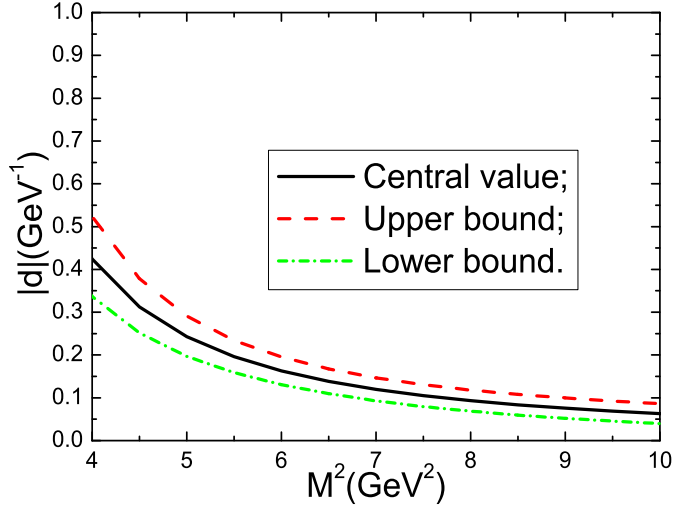}
     \caption{The  electromagnetic coupling constant  $d$ with the Borel parameter $M^2$. }
\end{figure}
\begin{figure} \centering
  \includegraphics[totalheight=8cm,width=10cm]{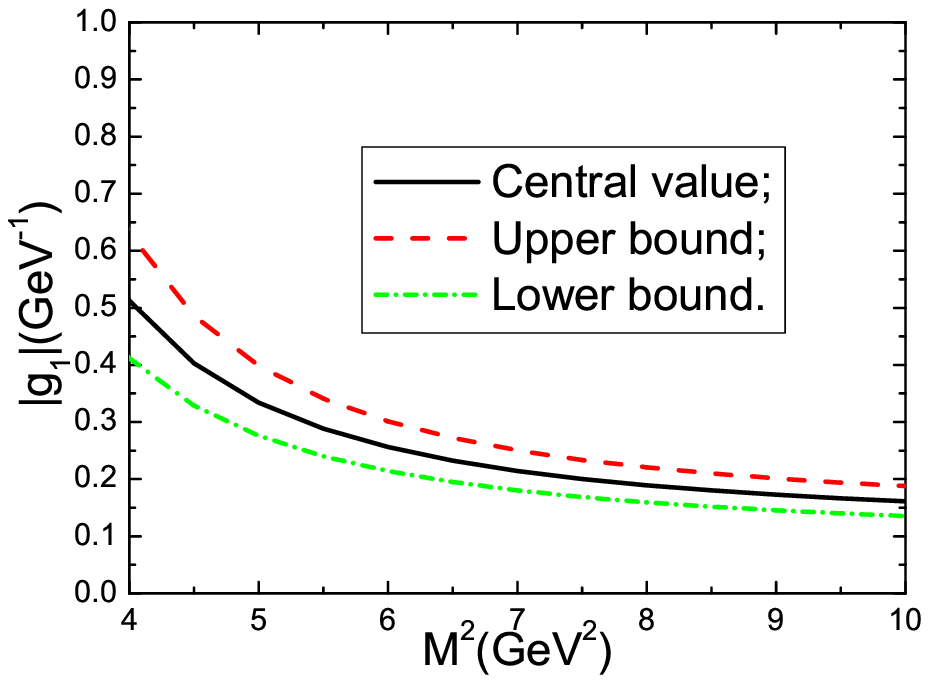}
     \caption{The electromagnetic coupling constant  $g_{1}$ with the Borel parameter $M^2$. }
\end{figure}
\begin{figure} \centering
  \includegraphics[totalheight=8cm,width=10cm]{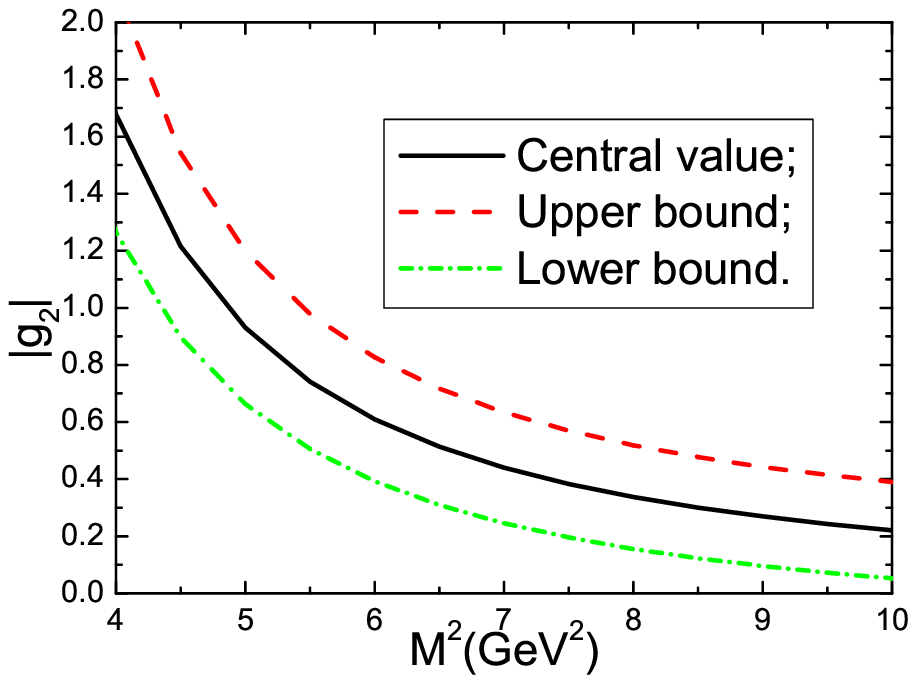}
     \caption{The electromagnetic coupling constant  $g_{2}$ with the Borel parameter $M^2$. }
\end{figure}
\begin{figure} \centering
  \includegraphics[totalheight=8cm,width=10cm]{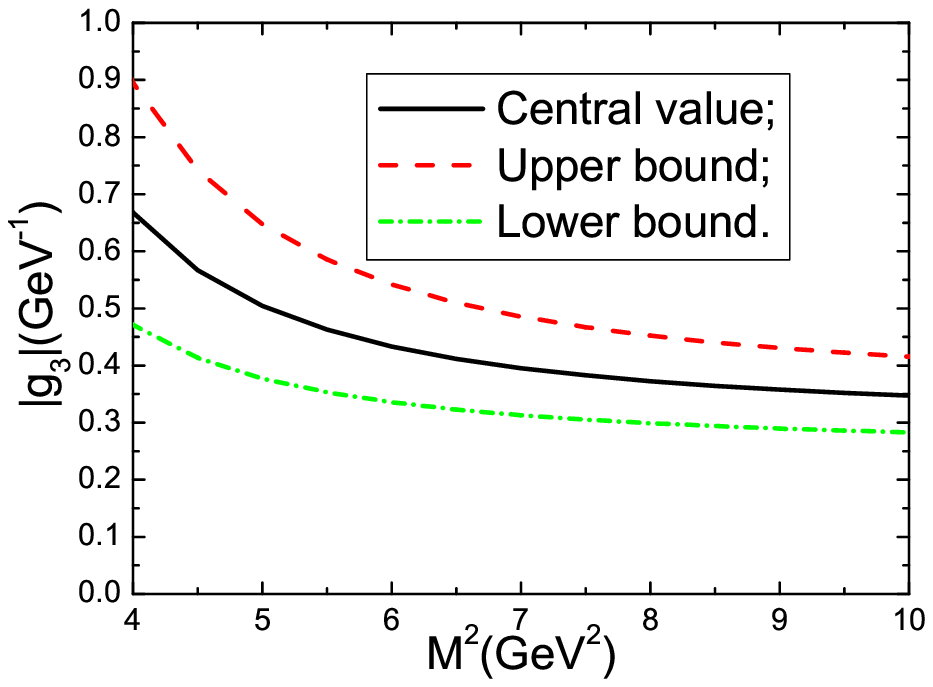}
     \caption{The electromagnetic coupling constant  $g_{3}$ with the Borel parameter $M^2$. }
\end{figure}
and  the radiative  decay widths,
\begin{eqnarray}
  \Gamma_{B_{s0}\rightarrow B_s^* \gamma} &=& \alpha d^2p^3=(1.3-13.6) \,\,\rm{KeV} \, ,\nonumber \\
  \Gamma_{B_{s1}\rightarrow B_s \gamma} &=& \frac{\alpha g_1^2p^3}{3}=(3.2-15.8) \,\,\rm{KeV} \, ,\nonumber \\
  \Gamma_{B_{s1}\rightarrow B_s^* \gamma} &=& \frac{\alpha g_2^2p^3(M_{B_{s1}}^2+M_{B_{s}^*}^2)}{3M_{B_{s1}}^2M_{B_{s}^*}^2}=(0.3-6.1) \,\,\rm{KeV} \, ,\nonumber \\
  \Gamma_{B_{s1}\rightarrow B_{s0} \gamma} &=&\frac{\alpha g_3^2p^3}{3}= (0.002-0.008) \,\,\rm{KeV} \,
  ,
 \end{eqnarray}
 where  $\alpha$ is the fine structure constant, and $p$ is the
 momentum of the final particles in the cental-of-mass frame of
  the initial meson. The values of the $p$ are about $0.28\,\,\rm{GeV}$, $0.34\,\,\rm{GeV}$, $0.30\,\,\rm{GeV}$ and
  $0.02\,\,\rm{GeV}$ for the radiative decays $B_{s0}\rightarrow B_s^*
  \gamma$, $B_{s1}\rightarrow B_s \gamma$, $B_{s1}\rightarrow B_s^*
  \gamma$ and $B_{s1}\rightarrow B_{s0} \gamma$ respectively. The
   decay widths  are proportional to $p^3$,
   the  decay $B_{s1}\rightarrow B_{s0} \gamma$ is kinematically suppressed and
  the  width $\Gamma_{B_{s1}\rightarrow B_{s0} \gamma}$ is
 rather small.

  The  energy gap between the $P$-wave and $S$-wave strange-charm mesons is larger  than the
  one  between two $P$-wave (or two $S$-wave) strange-charm mesons,
   $M_{D_{s1}}-M_{D_{s0}}=M_{D_{s}^*}-M_{D_s}=0.14 \,\,\rm{GeV}$
  and  $\frac{3M_{D_{s1}}+M_{D_{s0}}}{4}-\frac{3M_{D_{s}^*}+M_{D_s}}{4}=0.35\,\,\rm{GeV}$,
  the same relation  holds for their bottom cousins. The radiative decays
  between the $P$-wave and $S$-wave strange-bottom mesons are kinematically favorable
  comparing with  the
  internal transitions among the $P$-wave (or $S$-wave) mesons.
  There exists  a possibility that the radiative decay $B_{s0}\rightarrow B_{s1}
  \gamma$ can take place, its width is about  the same order
  of  the width $\Gamma_{B_{s1}\rightarrow B_{s0} \gamma}$, much smaller than the
   width $\Gamma_{B_{s0}\rightarrow B_s^*
  \gamma}$. It is not an ideal channel to search for the $B_{s1}$ or $B_{s0}$
  meson, we can search for the $B_{s1}$ and $B_{s0}$
  mesons in the invariant $B_s \gamma$ and $B^*_s \gamma$ mass distributions in the radiative
  decays at the LHCb.

In Ref.\cite{Faessler08}, Faessler et al take the $(0^+,1^+)$
doublet  $B_{s0}$ and $B_{s1}$   as the $BK$ and $B^*K$ molecules
respectively,  study the radiative decays,  and obtain the narrow
 widths  $\Gamma_{B_{s0}\rightarrow B_s^* \gamma} = 3.07 \,\,\rm{KeV}$ and  $\Gamma_{B_{s1}\rightarrow B_s \gamma} = 2.01
 \,\,\rm{KeV}$, which are much smaller than the  central values of the corresponding ones in the present work, see Eq.(12).
We can probe  the quark configurations of the mesons $B_{s0}$ and
$B_{s1}$ using their radiative decays. If the $(0^+,1^+)$ $D_s$ and
$B_s$ mesons  are  $Q\bar{s}$ cousins, the heavy quark symmetry
warrants that they have analogous decay hierarchy
$\Gamma_{B_{s1}\rightarrow B_s \gamma} \geq
\Gamma_{B_{s1}\rightarrow B_s^* \gamma} \geq
\Gamma_{B_{s1}\rightarrow B_{s0} \gamma}$ and
 $\Gamma_{D_{s1}\rightarrow D_s \gamma} \geq
\Gamma_{D_{s1}\rightarrow D_s^* \gamma} \geq
\Gamma_{D_{s1}\rightarrow D_{s0} \gamma}$ \cite{RadiativeD}. It is
indeed the case from the present analysis.  On the other hand, the
magnitudes are  quite different ($\Gamma_{D_{s1}\rightarrow D_s
\gamma} =(19-29)\,\rm{KeV}$, $\Gamma_{D_{s1}\rightarrow D_s^*
\gamma} =(0.6-1.1)\,\rm{KeV}$, $ \Gamma_{D_{s1}\rightarrow D_{s0}
\gamma}=(0.5-0.8\,\rm{KeV}$) due to the fact that the heavy quarks
$b$ and $c$ have different electric charge, the strange-bottom
mesons are neutral while the strange-charm mesons are charged. From
Eqs.(5-9), we can  see that the electromagnetic coupling constants
are proportional to $C_s e_s+C_Q e_Q$, where the $C_s$ and $C_Q$ are
formal notations, the heavy quark electric charge  $e_Q$ have
significant effects.

 The strong couplings of the $(0^+,1^+)$ $B_s$ mesons  to the
nearby  thresholds  may result in some tetraquark components,
whether the nucleon-like bound states or deuteron-like bound states
\cite{Wang08}, the  tetraquark components may lead to smaller
radiative decay widths than the corresponding pure $b\bar{s} $
states.

The  central values of the  masses of the  $(0^+,1^+)$
strange-bottom  mesons from the QCD sum rules  are below the
corresponding $BK$ and $B^*K$
 thresholds  respectively \cite{Wang0712}, the decays $B_{s0}\rightarrow BK$ and
$B_{s1}\rightarrow B^*K$ are kinematically  forbidden.   In the
previous works \cite{Wang0801}, we have calculated the  strong
coupling constants $g_{B_{s0} B_s \eta}$ and $g_{B_{s1} B^*_s \eta}$
using  the light-cone QCD sum rules, studied the strong isospin
violation decays $B_{s0}\rightarrow B_s\eta\rightarrow B_s\pi^0$ and
$B_{s1}\rightarrow B_s^*\eta\rightarrow
 B_s^*\pi^0$, and observed that the decay widths are about several $\rm{KeV}$ due to
the small $\eta-\pi^0$ transition matrix \cite{Dashen},
$\Gamma_{B_{s1} \to B_s^* \pi} = (5.3-20.7)\,\, \rm{KeV}$ and
$\Gamma_{B_{s0}\to B_s \pi} = (6.8-30.7)\,\, \rm{KeV}$.

There are two degenerate $P$-wave strange-bottom doublets: the
$j_q=\frac{1}{2}$
  states $B_{s0}$ and $B_{s1}$, and the $j_q=\frac{3}{2}$ states $B^*_{s1}$ and $B_{s2}^*$.
   If kinematically
allowed, the states with $j_q=\frac{1}{2}$ can decay via an $S$-wave
transition, while the $j_q=\frac{3}{2}$ states undergo a $D$-wave
transition; the decay widths of the states with $j_q=\frac{1}{2}$
are expected to be much broader than the corresponding
$j_q=\frac{3}{2}$ states. Our numerical results indicate the widths
of the $j_q=\frac{1}{2}$ states are also narrow
\cite{Wang0712,Wang0801}.

 We can search for the
$(0^+,1^+)$ strange-bottom  mesons $B_{s0}$ and $B_{s1}$ in the
invariant $B_s \pi^0$ and $B^*_s \pi^0$ mass distributions in the
strong decays or in the invariant $B_s^*\gamma$ and $B_s\gamma$ mass
distributions in the radiative decays. Those mesons can be observed
at the LHCb, where the $b\bar{b}$  pairs will be copiously produced
with the cross section about $500\,\mu b$ \cite{LHCbook}.

\section{Conclusion}
In this article, we   assume that the  $(0^+,1^+)$ strange-bottom
mesons $B_{s0}$ and $B_{s1}$ are the conventional $b\bar{s}$ mesons,
and calculate the electromagnetic coupling constants $d$, $g_1$,
$g_2$ and $g_3$ using the light-cone QCD sum rules. Then we study
the radiative decays $B_{s0}\rightarrow B_s^* \gamma$,
$B_{s1}\rightarrow B_s \gamma$, $B_{s1}\rightarrow B_s^* \gamma$ and
$B_{s1}\rightarrow B_{s0} \gamma$, and observe that the decay widths
are rather narrow. We can search for the  mesons $B_{s0}$ and
$B_{s1}$ in the invariant $B_s \pi^0$ and $B^*_s \pi^0$ mass
distributions in the strong decays or in the invariant $B_s^*\gamma$
and $B_s\gamma$ mass distributions in the radiative decays at the
LHCb.

\section*{Acknowledgments}
This  work is supported by National Natural Science Foundation,
Grant Number 10775051, and Program for New Century Excellent Talents
in University, Grant Number NCET-07-0282.

 \section*{Appendix}
 The   light-cone distribution amplitudes of the photon are  parameterized
as \cite{LCDA}
\begin{eqnarray}
\phi_\gamma(u)& = & 6 u \bar u \left[1+\varphi_2 C_2^{\frac{3}{2}}(2 u- 1) \right ] \, ,\nonumber \\
\mathbb{A}(u) &=& 40 u^2 \bar u^2 (3k - k^+ + 1) + 8 (\zeta_2^+ - 3\zeta_2)\big[u \bar u (2 + 13u \bar u) \nonumber\\
&&+ 2u^3(10 - 15u + 6u^2) \ln u + 2\bar u^3(10 - 15 \bar u + 6 \bar u^2)\ln \bar u \big]\, ,\nonumber \\
h_\gamma(u)&=&-10 (1 + 2  k^+) C_2^{\frac{1}{2}} (2 u - 1) \, ,\nonumber \\
\psi^v(u)&=& 5\left[3 (2  u - 1)^2 - 1\right] + \frac{3}{64} \left[15  \omega^V_\gamma - 5   \omega^A_\gamma \right] \left[3 - 30 (2  u - 1)^2 + 35 (2  u - 1)^4\right] \, ,\nonumber\\
\psi^a(u)&=&  \left[1-(2  u - 1)^2\right] \left[5(2 u -1)^2-1\right]  \frac{5}{2} \left[1+ \frac{9}{16} \omega^V_\gamma - \frac{3}{16}   \omega^A_\gamma \right]\, , \nonumber \\
{\cal V}(\alpha_q, \alpha_{\bar q}, \alpha_g) &=& 540 \omega_\gamma^V ( \alpha_q-\alpha_{\bar q}) \alpha_q \alpha_{\bar q} \alpha_g^2 \ , \nonumber\\
{\cal A}(\alpha_q, \alpha_{\bar q}, \alpha_g) &=& 360 \alpha_q
\alpha_{\bar q} \alpha_g^2  \left[ 1+ \omega_\gamma^A \frac{1}{2}(7
\alpha_g-3) \right] \, ,\nonumber \\
 {\cal S}(\alpha_q, \alpha_{\bar q}, \alpha_g) &=&  30 \alpha_g^2 \left\{(k + k^+)(1 - \alpha_g) + (\zeta_1 + \zeta_1^+)(1 - \alpha_g)(1 - 2 \alpha_g) \right.\nonumber \\
&&\left. + \zeta_2 [3 (\alpha_{\bar q} - \alpha_q)^2 - \alpha_g(1 - \alpha_g)]\right\} \, ,\nonumber \\
\tilde {\cal S}(\alpha_q, \alpha_{\bar q}, \alpha_g) &=&  - 30 \alpha_g^2 \left\{(k - k^+)(1 - \alpha_g) + (\zeta_1 - \zeta_1^+)(1 - \alpha_g)(1 - 2 \alpha_g) \right.\nonumber\\
&&\left. +\zeta_2[3(\alpha_{\bar q} - \alpha_q)^2 - \alpha_g(1 - \alpha_g)]\right\} \, , \nonumber \\
{\cal T}_1 (\alpha_q, \alpha_{\bar q}, \alpha_g) &=&  -120 (3 \zeta_2 + \zeta_2^+)(\alpha_{\bar q} - \alpha_q) \alpha_q \alpha_{\bar q} \alpha_g \, , \nonumber \\
{\cal T}_2 (\alpha_q, \alpha_{\bar q}, \alpha_g) &=& 30 \alpha_g^2 (\alpha_{\bar q} - \alpha_q)
\left[(k - k^+) + (\zeta_1 - \zeta_1^+)(1 - 2 \alpha_g) + \zeta_2(3 - 4 \alpha_g)\right] \, ,\nonumber \\
{\cal T}_3 (\alpha_q, \alpha_{\bar q}, \alpha_g) &=&-120 (3 \zeta_2 - \zeta_2^+)(\alpha_{\bar q} -
 \alpha_q)\alpha_q  \alpha_{\bar q} \alpha_g \, ,\nonumber \\
{\cal T}_4 (\alpha_q, \alpha_{\bar q}, \alpha_g) &=&  30 \alpha_g^2
(\alpha_{\bar q} - \alpha_q)\left[(k + k^+) + (\zeta_1 +
\zeta_1^+)(1 - 2 \alpha_g) + \zeta_2(3 - 4 \alpha_g)\right]  \, .
\end{eqnarray}


\begin{thebibliography}{99}


\bibitem{CDF}  T. Aaltonen et al, Phys. Rev. Lett. {\bf 100} (2008) 082001.

\bibitem{D0} V. Abazov et al, Phys. Rev. Lett. {\bf 100} (2008) 082002.

\bibitem{LHCbook} M. Kramer and  F. J. P. Soler, "Large hadron collider
phenomenology", Taylor $\&$ Francis, 2004.


\bibitem{BBmeson1} D. Ebert, V. O. Galkin and R. N. Faustov, Phys. Rev. {\bf D57} (1998)
5663.

\bibitem{BBmeson2} S. Godfrey and R. Kokoski, Phys. Rev. {\bf D43} (1991)
1679.

\bibitem{BBmeson3} W. A. Bardeen, E. J. Eichten and C. T. Hill, Phys. Rev. {\bf D68}
(2003) 054024.



\bibitem{BBmeson4} P. Colangelo, F. De Fazio and R. Ferrandes, Nucl. Phys. Proc. Suppl.
{\bf 163} (2007) 177.

\bibitem{BBmeson5} A. M. Green et al, Phys. Rev. {\bf D69}
(2004) 094505.

\bibitem{BBmeson6} M. Di Pierro and E. Eichten, Phys. Rev. {\bf D64} (2001)
114004.

\bibitem{BBmeson7} J. Vijande, A. Valcarce and F. Fernandez,
Phys. Rev. {\bf D77} (2008) 017501.

\bibitem{BBmeson8} M. A. Nowak, M. Rho and I. Zahed, Acta. Phys. Polon. {\bf B35}
(2004) 2377.

\bibitem{BBmeson9} I. W. Lee, T. Lee, D. P. Min and B. Y. Park, Eur. Phys. J. {\bf C49}
(2007) 737.

\bibitem{BBmeson10} I. W. Lee and T. Lee, Phys. Rev. {\bf D76} (2007) 014017.



\bibitem{Simonov07} A. M. Badalian, Yu. A. Simonov and M. A. Trusov,
Phys. Rev. {\bf D77} (2008) 074017.


\bibitem{Matsuki05} T. Matsuki, K. Mawatari, T. Morii  and K. Sudoh, Phys. Lett. {\bf B606} (2005) 329.

\bibitem{Matsuki07} T. Matsuki, T. Morii and K. Sudoh, Prog. Theor. Phys. {\bf 117} (2007) 1077.

\bibitem{Wang0712} Z. G. Wang,  Chin. Phys. Lett. {\bf 25} (2008) 3908.




\bibitem{Wang0801}  Z. G. Wang, Eur. Phys. J. {\bf C56} (2008) 181.



\bibitem{GI} S. Godfrey and N. Isgur, Phys. Rev. {\bf D32} (1985) 189.

\bibitem{Barik} N. Barik and P. C. Dash, Phys. Rev. {\bf D49} (1994) 299.

\bibitem{Jaus96} W. Jaus, Phys. Rev. {\bf D53} (1996) 1349.


\bibitem{Ebert2002} D. Ebert, R. N. Faustov, and V. O. Galkin, Phys. Lett. {\bf B537} (2002) 241.

\bibitem{Dosch96} H. G. Dosch and S. Narison, Phys. Lett. {B368} (1996) 163.

\bibitem{Aliev96} T. M. Aliev, D. A. Demir, E. Iltan, and N. K. Pak, Phys. Rev. {D54} (1996) 857.

\bibitem{ZHLi2002} Z. H. Li, X. Y. Wu and T. Huang, J. Phys. {\bf G28} (2002) 2583.


\bibitem{Cheng1993} H. Y. Cheng, C. Y. Cheung, G. L. Lin , Y. C. Lin, T. M. Yan
 and  H. L. Yu, Phys. Rev. {\bf D47} (1993) 1030.

\bibitem{Colangelo1993} P. Colangelo, F. De Fazio and G. Nardulli,
Phys. Lett. {\bf B316} (1993) 555.


\bibitem{Amu1992} J. F. Amundson, C. G. Boyd, E. E. Jenkins, M. E. Luke, A. V. Manohar,
J. L. Rosner, M. J. Savage and M. B. Wise, Phys. Lett. {\bf B296}
(1992) 415.

\bibitem{Choi2007} H. M. Choi, Phys. Rev. {\bf D75} (2007) 073016.


\bibitem{HQET1997}  R. Casalbuoni, A. Deandrea, N. Di Bartolomeo, F. Feruglio, R. Gatto and G.
Nardulli, Phys. Rept. {\bf 281} (1997) 145.

\bibitem{HQPbook} A. V. Manohar and M. B. Wise, Camb. Monogr. Part. Phys. Nucl. Phys. Cosmol. {\bf 10} (2000) 1.




\bibitem{Godfrey2003PLB} S. Godfrey,  Phys. Lett.  {\bf B568} (2003) 254.


\bibitem{Liu2006CQM} X. Liu, Y. M. Yu, S. M. Zhao and  X. Q. Li, Eur. Phys. J. {\bf C47} (2006) 445.


\bibitem{Godfrey2005}  S. Godfrey, Phys. Rev. {\bf D72} (2005)
054029.

\bibitem{Close2005CQM}  F. E. Close and E. S. Swanson, Phys. Rev. {\bf D72} (2005)
094004.

\bibitem{Colangelo2003PLB} P. Colangelo and F. De Fazio, Phys. Lett.  {\bf B570} (2003) 180.

\bibitem{Colangelo2004MPLA} P. Colangelo, F. De Fazio and R. Ferrandes, Mod. Phys. Lett.
{\bf A19} (2004) 2083.

\bibitem{Mehen2004} T. Mehen and  R. P. Springer, Phys. Rev. {\bf D70} (2004) 074014.



\bibitem{RadiativeD} P. Colangelo, F. De Fazio and A. Ozpineci, Phys. Rev. {\bf D72} (2005) 074004.


\bibitem{Wang2007R}   Z. G. Wang, Phys. Rev. {\bf D75} (2007) 034013.


\bibitem{Gamermann2007} D. Gamermann, L. R. Dai and E. Oset, Phys. Rev. {\bf C76} (2007) 055205.


\bibitem{PDG} C. Amsler et al, Phys. Lett. {\bf  B667} (2008) 1.

\bibitem{Wang08} Z. G. Wang, Phys. Rev. {\bf D77} (2008) 054024.

\bibitem{LCSR1} I. I. Balitsky, V. M. Braun and A. V. Kolesnichenko, Nucl. Phys.
{\bf B312} (1989) 509.

\bibitem{LCSR2} V. L. Chernyak and I. R. Zhitnitsky, Nucl. Phys. {\bf B345} (1990)
137.

\bibitem{LCSR3} V. L. Chernyak and A. R. Zhitnitsky, Phys. Rept. {\bf 112} (1984)
173.

\bibitem{LCSR4} V. M. Braun and I. E. Filyanov, Z. Phys.  {\bf C44} (1989)
157.

\bibitem{LCSR5} V. M. Braun and I. E. Filyanov, Z. Phys. {\bf C48} (1990) 239.

\bibitem{LCSRreview} P. Colangelo and A. Khodjamirian, hep-ph/0010175.


\bibitem{SVZ791} M. A. Shifman, A. I. Vainshtein and V. I. Zakharov,
Nucl. Phys. {\bf B147} (1979) 385.

\bibitem{SVZ792} M. A. Shifman, A. I. Vainshtein and V. I. Zakharov,
Nucl. Phys. {\bf B147} (1979)  448.

\bibitem{Reinders85} L. J. Reinders, H. Rubinstein and S. Yazaki, Phys. Rept. {\bf 127} (1985) 1.

\bibitem{LCDA} P. Ball, V. M. Braun and N. Kivel, Nucl. Phys. {\bf B649} (2003) 263.


\bibitem{Belyaev94} V. M. Belyaev, V. M. Braun, A. Khodjamirian and R. R\"uckl, Phys.
Rev. {\bf D51} (1995) 6177.



\bibitem{WangD1} Z. G. Wang and S. L. Wan, Phys. Rev. {\bf D73} (2006)
094020.

\bibitem{WangD2}Z. G. Wang, J. Phys. {\bf G34} (2007) 753.

\bibitem{Wang06} Z. G. Wang and S. L. Wan, Phys. Rev. {\bf D74} (2006) 014017.


\bibitem{Guo0712} F. K. Guo, S. Krewald and Ulf-G. Meissner, Phys. Lett. {\bf B665} (2008) 157.



\bibitem{BaBar0607} B. Aubert et al, Phys. Rev. Lett. {\bf 97} (2006) 222001.

\bibitem{2Ds-1} E. van Beveren and G. Rupp, Phys. Rev. Lett. {\bf 97} (2006) 202001.

\bibitem{2Ds-2} F. E. Close, C. E. Thomas, O. Lakhina and E. S. Swanson, Phys. Lett. {\bf B647 } (2007) 159.






\bibitem{HDress1}  N. A. Tornqvist, Z. Phys. {\bf C68} (1995) 647.

\bibitem{HDress2} E. van Beveren and  G. Rupp, Phys. Rev. Lett. {\bf 91} (2003)
012003.

\bibitem{HDress3} Yu. A. Simonov and J. A. Tjon, Phys. Rev. {\bf D70} (2004) 114013.



\bibitem{Wang04BS} Z. G. Wang, W. M. Yang and S. L. Wan, Nucl. Phys. {\bf A744} (2004)
156.

\bibitem{Verde-Velasco07} J. M. Verde-Velasco, arXiv:0710.1790; and references  therein.






\bibitem{Faessler08}  A. Faessler, T. Gutsche, V. E. Lyubovitskij and Y. L.
Ma, Phys. Rev. {\bf D77} (2008) 114013.

\bibitem{Dashen} R. F. Dashen, Phys. Rev. {\bf 183} (1969) 1245.



\end{thebibliography}
\end{document}